# The Dynamics of a Mobile Phone Network

*Cesar A. Hidalgo[1*], C. Rodriguez-Sickert [2,]*

[1] *Center for Complex Network Research, Department of Physics, University of Notre Dame.*

[2] *Department of Sociology, Pontificia Universidad Catolica de Chile.*

## Abstract

*The empirical study of network dynamics has been limited by the lack of longitudinal data. Here we introduce a quantitative indicator of link persistence to explore the correlations between the structure of a mobile phone network and the persistence of its links. We show that persistent links tend to be reciprocal and are more common for people with low degree and high clustering. We study the redundancy of the associations between persistence, degree, clustering and reciprocity and show that reciprocity is the strongest predictor of tie persistence. The method presented can be easily adapted to characterize the dynamics of other networks and can be used to identify the links that are most likely to survive in the future.*

---

[*] to whom correspondence should be addressed: chidalgo@nd.edu

## Introduction

Physicists are no strangers to the study of social networks. During the last decade several groups have studied the structure of social networks as expressed in e-mails [1,2], cellular phones [3,4] and work relationships such as starring on a movie [5] or collaborating in a paper [6]. Studying the dynamics of such social systems however, has been limited by the lack of longitudinal data and as a result only a few studies on the dynamics of interpersonal connections have been produced [1,3,7].

In principle there are many factors that could affect the stability of a social link [8,9,10]. The aim of this paper is not to determine such factors, but to study the coupling between the structure of the network as characterized in previous studies [11,12] and the temporal stability of its links.

Here we use a years worth of mobile phone data as a proxy for the structure and dynamics of a social network involving close to two million people. Automatically collected communication records have been proposed as a source of reliable data about personal connections [13]. Email data has been used to study social processes such as tie formation [1] and social structure [2]. The citations patterns of web logs have been used to study the spread of political opinions [14] and the growth of an online dating community has been documented [7]. Communication records overcome problems of survey data such as subjective biases on the respondents and the intrinsic limitations of ego-centered networks, like their unreliability measuring degree and clustering.

It is not our intention to claim that cellular phone communications fully capture social exchange. A social network is expressed through a host of interactions ranging from emails to sexual contacts. Cellular phone calls are just one of the ways in which a social tie is expressed. People in close social contact tend to express ties through multiple interaction channels [15], such as email, cell phone communications, instant messaging and face to face interaction. However, there are arguments favoring the use of cellular phone calls as a relevant proxy for large-scale social networks. Specifically, it has been shown that objective measures as the one we use in our study can accurately predict self reported friendships [16]. The interest of the community has been expressed through the recent emergence of a literature on mobile phone networks in which using this and other data sets people have studied the strength of social ties in cross sections of the network [4] and the dynamics of social groups [1,3].

There are also some technical aspects that favor the use of a mobile phone network as a proxy for social interactions. Mobile phone numbers are unlisted, thus knowing them reveals some sort of social connection between caller and callee. Also, cellular phones were the most widespread information technology at the time this data was collected; with a penetration larger than 40% worldwide and close to 100% in developed countries, like the one considered in this study. During the same time period, internet penetration was just over 13% worldwide and 51% for developed countries (MDGS indicators U.N. http://mdgs.un.org/unsd/mdg) making it the most complete method to study social interactions on the population scale. In addition, mobile phone usage has been particularly democratic to the extent that it has homogeneously penetrated different social strata [17]

## Data

Our study is based on mobile phone calls. The data consists of 7,948,890 voice calls between 1,950,426 users of a service provider holding approximately 25% of an industrialized country's market. The data consist of ten panels collected between April 15, 2004 and March 31, 2005. Each panel summarizes 15 days of mobile phone calls between the members serviced by the provider who facilitated the data. Not every panel is available, as this was the way in which data was made available to us. We consider only agents that made or received at least one call in each panel to avoid dealing with dropouts or new subscribers. We hereafter assume that at high service penetration levels (~100%) agents serviced by a particular provider are equivalent to a random sample. In our network nodes are mobile phones, which we interpret as people and links are the calls connecting them.

## *Results*

### The Persistence of Ties

We measure the stability of ties across time as the number of panels in which a link is observed, over the total number of panels. We denote this measure as persistence which can be expressed as:

$$P_{ij} = \frac{\sum_T A_{ij}(T)}{M}, \qquad (1)$$

where $A_{ij}(T)$ is 1 if nodes $i$ and $j$ communicated on panel $T$ and 0 otherwise, whereas $M$ is the total number of panels. Persistence is the probability of observing a tie when observing a panel of network data. Our definition of persistence has a resolution that depends on the duration of the panels. If we consider panels with a duration comparable to the one of links, (~ minutes in the case of phone calls), our definition of persistence just gives the number of times a tie appeared. Whereas when we consider panels lasting considerably longer than the typical duration of a link, our definition of persistence will capture the stability of a link on a larger, coarse-grained temporal scale. Our data set consists of 10 panels, each summarizing 15 days of voice call activity. Thus we measure persistence on a monthly to yearly time scale.

We illustrate our definition of persistence using four different panels of a five node network (Fig. 1 a). In this example, the link between nodes 2 and 4 is present in all panels while the one between nodes 1 and 2 is present only in half of them. We say that the persistence of the link between nodes 2 and 4 is 4/4 while the persistence of the link connecting nodes 1 and 2 is 2/4. Each panel gives a binary representation of the network, where a link is either present or not. Our definition of persistence summarizes the dynamics of all binary panels by assigning a weight to each link. Thus persistence is a change of representation that allows us to study many panels as a single weighted network (Fig. 1 b).

Our measure of persistence weakly increases with the number of times a link is observed, persistence indicates stability, as understood in previous studies [18,19]. However, given that we measure whether the link is observed in $N>2$ panels, it will not describe a link dichotomously as stable or unstable, but will give the degree of stability $1/N \leq P \leq 1$, rewarding those links expressed consistently in many panels.

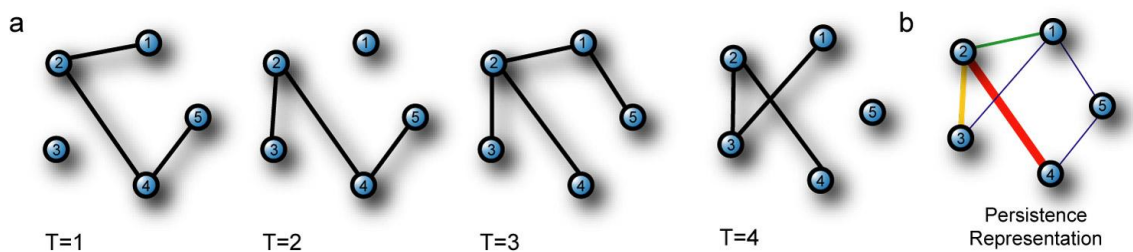

**Fig 1** Definition of Persistence. **a** Four panels of a five node network in which not all links are equally persistent. **b** Persistence representation of the four panels presented in **a**.

Persistence is a tie attribute that can be defined for a particular node as the average persistence of all its ties. We denote this as *perseverence* and define it as

$$P_i = \frac{1}{k_i} \sum_j P_{ij}, \qquad (2)$$

where $k_i$ is the degree, or number of connections of the $i^{th}$ node. We will use this quantity to study what characterizes nodes carrying persistent ties.

Our definition has limitations. One could claim we are unfairly punishing newly formed links. An alternative strategy would be to consider only the links involved in the first panel; however an exercise in this line showed us that there is a strong selection bias towards stable links when we consider such an option. For example, links appearing only once, on the second to tenth panel, will not be considered if we set our benchmark on the first panel only. Our definition also does not differentiate between links active half of the time or those active during a particular half of the year. We do not propose our measure as the ultimate way to reduce a set of network panels into a weighted network, but as a simple way to do so, allowing us to characterize to first approximation the stability of a network's links.

## *Results*

### Global Analysis of the Persistence of Ties

Figure 2a shows the persistence histogram for the voice call network. The distribution is bimodal, meaning that ties tend to be either active most of time or rarely expressed. This is known as a core-periphery structure [13], where stable ties compose a person's social core and unstable ties connect people to the more peripheral actors in their life.

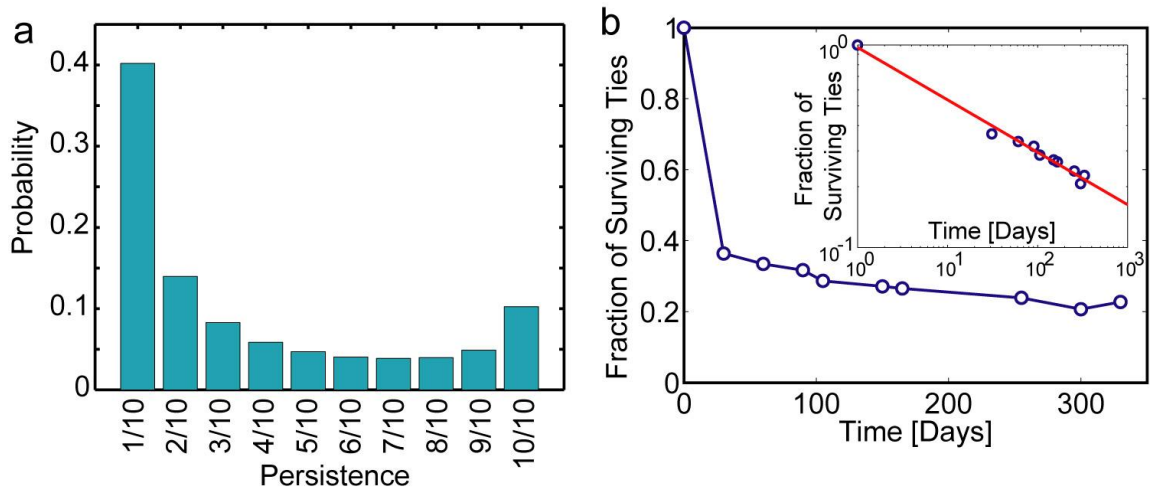

**Fig 2**. Persistence across a cellular phone network **a** Distribution of persistence for all links **b** Fraction of surviving ties as a function of time. The inset shows the same plot in a double logarithmic scale. The continuous line is $t^{-1/4}$

The decay of ties as a function of time can be approximated by a power-function (Fig. 2 b), in agreement with the 4-year study performed by Burt [20]. The fact that the survival probability of a tie can be approximated by $\sim t^{-\alpha}$ with $\alpha = 0.25 \pm 0.07$ indicates that a great number of ties disappear quickly, while others tend to stay for very long periods of time. On average, less than 40% of the ties are conserved after 15 days. After this initial drop however, ties disappear slowly and more than 20% of the ties remain after a year. We note that the discreteness of our data does not allow us to prove that tie decay follows a power-law, yet the graphic analysis can be considered as suggestive evidence motivating a hypothesis in this direction.

### Network Structure and the Persistence of Ties

*Bivariate Analysis*

Figure 3 a show a fragment of the mobile call network extracted by considering all connections up to 3 links away of a randomly chosen user. Although this example shows less than the 0.0008% of our network, it visually summarizes the correlations between persistence, perseverance and the topological attributes of the mobile call network. In particular, we find that these temporal attributes correlate with topological variables such as the number of connection or degree $k_i$, the average reciprocity of a node $r$ (fraction of ties containing both, incoming and outgoing calls) and the clustering coefficient of a node $C_i$ defined as:

$$C_i = \frac{2\Delta}{k_i(k_i-1)} , \qquad (3)$$

where $\Delta$ is the number of triads in which the node is involved. Figure 3b shows a histogram of persistence split into 9 different degree categories revealing that persistent links represent a large fraction of the connections for low degree nodes while transient links are more common for large degree nodes. However, the number of persistent ties grows as a function of degree, meaning that although on average the persistence of high degree nodes is lower, in absolute terms their core is larger.

Figure 3d shows the distribution of persistence divided by clustering coefficient categories, indicating that highly clustered nodes tend to have relatively large cores. In the core periphery context, this means that persevering nodes are located in dense parts of the social network (Fig. 3a I) while those in sparser parts tend to have non-persistent ties acting as bridges which interruptedly connect different parts of the network (Fig. 3a II). Finally, we split the distribution of persistence by reciprocity (figure 3e) and observe that nodes with more reciprocated ties tend to be more persistent.

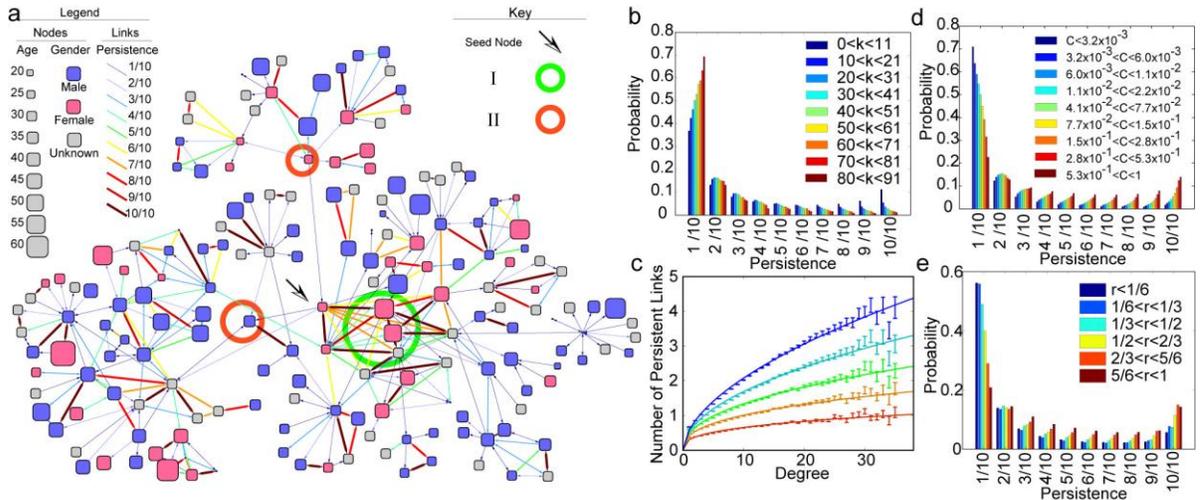

**Fig 3** Network structure and the persistence of ties **a** A fragment of the network extracted by considering up to the second neighbor of a randomly chosen node (indicated by a black arrow). **b** Distribution of persistence divided into nine degree categories. **c** Number of persistent links defined as those with a persistence of, from top to bottom: 5/10, 6/10, 7/10, 8/10, 9/10 and 10/10. **d** Distribution of persistence divided into nine clustering categories. **e** Distribution of persistence divided into five different reciprocity segments.

*Multivariate Analysis*

In the previous section we presented a bivariate analysis in which we analyzed the effect of three single structural variables and found that persistence depends monotonically on all of them (degree, clustering coefficient and reciprocity). The observed correlations however, might well be redundant. To check if this is the case we perform a multivariate analysis to quantify the effect of each of these variables on the persistence of ties. Because of the large number of observations considered (~ 2 million nodes, ~8 million ties) the confidence intervals of the regressions do not spread far from the predicted values. Hence we concentrate our discussion on the relative magnitude of the effects rather than on their significance.

On a social network, it is a well known fact that agents tend to connect to others of similar degree significantly more than random [21,22]. It is not known however, whether links connecting same degree agents tend to be more stable than those connecting different degree agents. To study this effect we performed a regression in which we study the persistence of a link as function of the difference in degree of the nodes that link connects. Furthermore, we also include in the regression the difference in clustering and average reciprocity of nodes connected by a particular link. In addition to this, we consider two link attributes, the reciprocity of links *R*, was there ever a panel in which caller and callee reciprocally called each other, and the topological overlap associated with that link which is defined as

$$T.O.(i, j) = \sqrt{\frac{O_{ij}^2}{k_i k_j}} \quad , \tag{4}$$

Where $O_{ij}$ is the number of neighbors that agents *i* and *j* have in common and $k_i$ and $k_j$ are their respective degrees.

Together these 5 variables explain 40% of the variance in persistence (Table 1 $R^2$ = 0.397). The contribution of each one of them can be isolated by considering the partial regression coefficients [23], which are a way to quantify how much of the variance is explained by each one of the covariates used in a regression. This technique shows that assortative mixing is not associated with the persistence of ties. Whereas the reciprocity of the links (0 non-reciprocal, 1 reciprocal) explains 26% of persistence followed by topological overlap which explains 3.4 % of the variance in persistence.

| Pearson's Correlation | ΔC | Δk | Δr | R | TO | Persistence |
|---|---|---|---|---|---|---|
| ΔC | 1 | 0.023 | 0.15 | 0.11 | 0.23 | 0.15 |
| Δk |  | 1 | 0.02 | -0.13 | -0.19 | -0.16 |
| Δr |  |  | 1 | -0.68 | -0.073 | 0.033 |
| R |  |  |  | 1 | 0.2964 | 0.5886 |
| TO |  |  |  |  | 1 | 0.3537 |
| Regression Coefficients | 0.09 | 0.002 | 0.15 | 0.35 | 0.56 |  |
| Partial Correlations | 0.0027 | 0.0032 | 0.007 | 0.26 | 0.034 |  |

**Table 1** Persistence of ties and link attributes

In the previous section we showed that high degree agents had on average less persistent ties than low degree agents. We also saw that highly clustered agents tended to have a larger number of persistent connections and that reciprocal ties tend to be more persistent in average. Again, we explore the redundancy of such statements using linear regression and split the contribution to perseverance from each of these variables by calculating their partial correlations (Table 2). Together, these variables explain almost 50% of the variance in perseverance ($R^2=0.49$). Their contributions however are quite uneven. When we look at the partial correlation coefficients extracted from our linear model we find that most correlations vanish and the biggest contribution to perseverance is given by the average reciprocity r of an agents ties, which explains *27%* of the variance. The negative effect of degree of the persistence of an agents ties is still present, but greatly ameliorated. This means that high degree agents which reciprocate their ties have more persistent ties as well. The negative effect of an agents degree on the persistence of its ties is in large part explained by the fact that high degree agents tend to reciprocate less of their ties. Similarly, the clustering coefficient *C*, which appeared as the strongest predictor in the bi-variate case, explains only 6% of the variance when reciprocity and degree are taken into account. This shows that cliques are formed by reciprocal ties minimizing the additional information about persistence carried by cliques themselves.

| Pearson's Correlation | C | K | r | Perseverance |
|---|---|---|---|---|
| C | 1 | -0.51 | 0.49 | 0.64 |
| k |  | 1 | -0.34 | -0.45 |
| R |  |  | 1 | 0.62 |
| Regression Coefficients | 0.0598 | -0.0122 | 0.3626 |  |
| Partial Correlation | 0.062 | 0.11 | 0.27 |  |

**Table 2** Correlations and regressions between node attributes and perseverance

## Predictability: Using topology to infer future ties.

How well can we predict the stability of ties starting from a single panel? As mentioned before, persistence is a time-like, vertical variable and is not constrained to correlate with space-like, horizontal variables. As we saw from our multivariate analysis, structural variables can be redundant [24, 25] and thus it is important to take into account their correlations to unveil their real contribution to the persistence of ties. Can we use this information to predict which ties persist in time? To answer this question we looked at our first data panel and used different criteria to predict which ties will be stable. We then looked at the fraction of these ties appearing after 1, 3, 6, 9 and 12 months and gauged the accuracy of our predictions by measuring their Positive Predictive Value (*PPV*) defined as:

$$PPV = \frac{TP}{TP+FP}, \quad (5)$$

where *TP* is the number of true positives and *FP* is the number of false positives.

We begin by testing the prediction that all ties observed as reciprocal in the first panel will be conserved in the future. For this hypothesis, the *PPV* ranges from 70% after one month to 43% after a year (Fig 4a). For comparison, we picked a random set of ties and found a *PPV* of 35% after a month and 20% after a year.

We can improve our predictive power by using a more stringent criterion. If we consider all reciprocal links that also have a topological overlap larger than $TO \geq 0.01$ we improve the *PPV* of our prediction by 5%, while an even more stringent criterion based on a $TO \geq 0.1$, gives us an extra percent that allows us to predict with a *PPV* larger than 50% after one year.

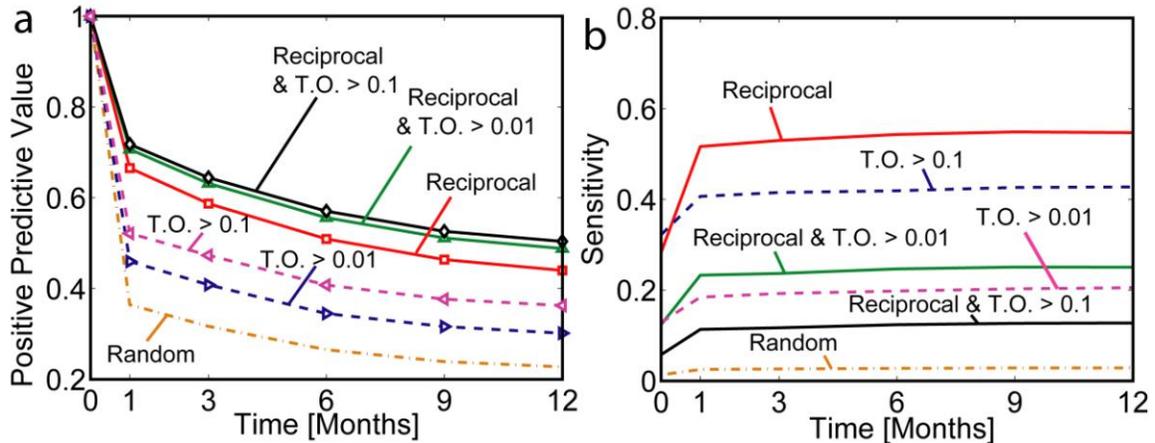

**Fig 4** Predicting future ties **a** Accuracy of tie prediction by randomly choosing ties (orange), choosing reciprocal ties (red), reciprocal ties with a *T.O.>0.01* (green), reciprocal ties with a *T.O.>0.01*, ties with a *T.O.>0.01* (blue) and a *T.O>0.1* (purple). **b** Sensitivity of the predictive methods presented in a. using the same color scheme.

The increase in accuracy brought by more stringent criteria reduces the number of links predicted to be persistent. Thus the sensitivity defined as:

$$S = \frac{TP}{TP + FN} \qquad (6)$$

where *FN* is the number of false negatives, decreases with the stringency of the criteria used but increases with time (Fig. 4 b). Hence there is a tradeoff between the accuracy of our prediction and the number of predictions we can make. Using the simple method presented above, an increase on accuracy comes with a decrease in sensitivity so more accurate predictions can be made only if we accept a reduction in the number of predictions being made.

Reciprocity appears to be the best predictor of persistence; however, it is not the only one. The fact that the variance explained by other structural variables was redundant with that explained by reciprocity allows us to use them as alternative predictor of ties. Figure 4a also shows the *PPV* obtained when we use topological overlap as our only predictive criterion. In this case we see that although the accuracy is lower, it is still significantly better than random. Thus the redundancy observed in the system can be turned into a predictive advantage and in the absence of information about the reciprocity of links we can use redundant measures to make good educated guesses about the existence of future ties.

## *Discussion*

We have defined and measured the persistence of ties in a one year period using 10 panels of data summarizing the activity of all voice calls carried by a mobile phone carrier from an industrialized country. We showed that the persistence of ties and perseverance of nodes depend on topological variables (degree, clustering, reciprocity and topological overlap). In our study, topological variables explain almost half of the variance in persistence. The stability of social ties is likely a behavioral attribute, thus it is not surprising that the local structure of the social network, that it is likely also a result of social behavior, predicts the persistence of ties.

Social connections ultimately affect processes such as collective decision [26,27] and coordinated consumption [28]. But not all social connections are equally important, some ties are stronger than others [29]. The strength of a social tie is not an absolute measure; hence there is a need to quantify the strength of ties using ad-hoc measures. Persistence is a way to quantify the temporal stability of ties, and therefore their strength, in one of the many possible dimensions that tie strength can be quantified. As longitudinal data becomes available, methods like the one introduced here can be used to quantify the strength of links and ultimately determine its effects on network dynamics.

The relationships shown here demonstrate that the temporal dynamics of social interactions are intrinsically coupled to the social network structure in such a way that the existence of a tie can be predicted, with a respectable accuracy, using a simple criterion.

Reciprocity is the stronger predictor of tie stability. If you do not want to loose that friend, you should better call him back once in a while.

## Acknowledgments

C.A. Hidalgo was partly supported by the Kellogg Institute at Notre Dame and acknowledges support from NSF grant ITR DMR-0426737, IIS-0513650 and the James S. McDonnell Foundation 220020084. C. Rodriguez-Sickert acknowledges Sam Bowles and the S.F.I. We thank Nicole Leete for proof reading our manuscript. Special acknowledgments to A.-L. Barabasi for providing the source data and discussing the manuscript.